\documentclass[aps,prl]{revtex4}
\newcommand{\ket}[1]{$|#1\rangle$}

\begin{document}
\title{Going beyond GHZ paradox}
\author{Dagomir Kaszlikowski$^{1, 2}$,
Darwin Gosal$^1$, E.J. Ling$^1$, \\ L.C. Kwek$^{1, 3}$, Marek \.Zukowski$^4$, and
C.H. Oh$^{1}$}

\address{
$^1$Department of Physics, National University of Singapore, 10 Kent Ridge Crescent, Singapore 119260 \\
$^2$Instytut Fizyki Do\'swiadczalnej, Uniwersytet Gda\'nski, PL-80-952, Gda\'nsk, Poland,\\
$^3$National Institute of Education, Nanyang Technological University, 1 Nanyang Walk, Singapore 639798\\
$^4$Instytut Fizyki Teoretycznej i Astrofizyki, Uniwersytet Gda\'nski, PL-80-952, Gda\'nsk, Poland.}

\begin{abstract}
 We present  numerical data showing,
that three qutrit correlations for a pure state, which is not
maximally entangled, violate local realism more strongly than
three-qubit correlations. The strength of violation is measured
by the minimal amount of noise that must be admixed to the system
so that the noisy correlations have a local and realistic model.
\end{abstract}

\maketitle

The seminal paper of Greenberger, Horne, and Zeilinger \cite{GHZ}
has initiated a completely new phase in the discussions regarding
the Bell theorem \cite{BELL}.  Einstein-Podolsky-Rosen \cite{EPR} elements of
reality were suddenly ridiculed by a straightforward
argumentation. The physics community immediately noticed that the
increasing complexity of entangled systems does not lead to a less
pronounced disagreement with the classical views, but just the
opposite! Moreover, the disagreement exponentially grew with the
number of qubits involved in the GHZ-type entangled states.
Indeed, prior to the publication of Ref. \cite{GHZ}, it was
commonly perceived that everything regarding the Bell theorem is
known. However, the new insight has renewed the interest in the
Bell theorem and its implications.

Another widely shared perspective was that one cannot gain
additional useful insight into the Bell Theorem by increasing the
dimensionality of the entangled systems. Some papers even
suggested that in $N$ dimensional systems, increasing the
dimension $N$ effectively  brings the system closer and closer to
the classical realm. However, due to the fact that the $N
> 2$ dimensional systems can reveal the Kochen-Specker paradox
\cite{KOCHEN-SPECKER}, this view could be challenged. The advent
of the quantum information theory created the awareness that such
systems require much less entanglement to be non-separable than
qubits \cite{SEP}. Certain strange features like bound
entanglement \cite{HORODECKI} or inextensible product bases
\cite{BENNET}, suddenly emerged.

Recently, it was shown that  higher dimensional entangled systems
indeed may lead to stronger violations of local realism, even in
straightforward experimental situations involving only the von
Neumann-type experiments (with no sequential measurements, etc).

In the early nineties, the blueprints for straightforward Bell
tests involving higher dimensional systems were given (for a
summary see \cite{ZZH}). The idea was to use unbiased multiport
beam splitters to define the local observables. Surprisingly, it
turned out, that such observables suffice to reveal the fact that
pair of entangled higher dimensional systems violate local realism
more strongly than qubits \cite{KASZLIKOWSKI-PRL-2000}. This
result was obtained numerically by employing the  linear
optimization procedures to  search for underlying local realistic
joint probability distribution that would reproduce the quantum
prediction (with some noise admixture). The results were confirmed
analytically in Refs. \cite{CHQUTRIT} and \cite{COLLINS}. Later
 in Ref.
\cite{ACIN-GISIN}, it was shown that in the case of pairs of
entangled higher-dimensional systems, violations of local realism
are even stronger for {\em non-maximally} entangled states. In a
parallel research, it has been shown that higher dimensional
systems can lead to the GHZ-like paradox without inequalities
\cite{GHZKASZLIKOWSKI}, \cite{MASSAR}.

In view of all these facts, it is tempting to test the strength of
violation of local realism by triples of higher dimensional
systems (starting of course with three qutrits), and that for
non-maximally entangled states.

Since Bell-type inequalities for three qutrit systems are unknown
at the moment, it is necessary to invoke the numerical algorithm
first presented in \cite{BATURO}. As we shall see, some surprising
results can be obtained in this way.

We show below the result of our numerical analysis. It turns out that
\begin{enumerate}
\item There is a strong violation of local realism
(for the standard von-Neumann type measurements) for three qutrit
systems in the maximally entangled state, however it is not as
strong as in the case of the three entangled qubits.
\item Allowing non-maximally entangled states, the situation changes.
We find the three qutrit state which reveals correlations  much
much more resistant to noise, than those for entangled three
qubits (maximally entangled three qubit states give maximal
violation of local realism \cite{WERNER}, \cite{BRUKNER},
\cite{SCARANI}).
\end{enumerate}

%\section{Linear optimization}
In our numerical analysis, we consider a class of pure states of
three qutrits in the form of
\begin{equation}
|\psi\rangle = \sum_{g,i,j = 1}^{3} d_{gij}|g\rangle |i\rangle
|j\rangle \label{STATE}
\end{equation}
with {\em real} coefficients $d_{gij}$. The kets $|g\rangle,
|i\rangle, |j\rangle$ denote the orthonormal basis states for the
first, second and the third qutrit respectively. Three spatially
separated observers, Alice, Bob and Cecil, are allowed to perform
the measurement of two alternative local noncommuting trichotomic
observables on the state $|\psi\rangle$. We assume that they
measure observables defined by unbiased symmetric three-port
beamsplitters \cite{ZZH}. In such a situation the kets in
(\ref{STATE}) represent spatial beams, in which the particles can
propagate.
 The observers select the specific local
observables by setting appropriate phase shifts in the beams
leading to the entry ports of the beamsplitters. The overall
unitary transformation performed by such a device is given by
\begin{equation}
U_{j'j}={1\over\sqrt 3}\exp({i2\pi\over 3}j'j)\exp(\phi_{j}),
\label{UNITARY}
\end{equation}
 where $j$ denotes an input beam to the device, and $j'$ an
 output one, and
$\phi_j$ are the three phases that can be set by the local
observer (for a more detailed description see \cite{ZZH}). Please
note, that the actual physics of the device is irrelevant for our
theoretical discussion here, thus it suffices just to assume that
the observers perform their von Neumann measurements in the basis
which is related to the ``computational" basis of the initial
state (\ref{STATE}) by the transformation (\ref{UNITARY}). It is
interesting that the unitary transformation for all phase
settings leads to a new basis for the local qutrit, which is
unbiased with the respect to the ``computational" one.

Let us
denote Alice's local unitary transformations associated with her device
by $U_A(\vec{\phi}_0), U_A(\vec{\phi}_1)$,
Bob's by $U_B(\vec{\chi}_0), U_B(\vec{\chi}_1)$ and Cecil's by
$U_C(\vec{\delta}_0),U_C(\vec{\delta}_1)$, where the three component
vectors $\vec{\phi}_k,\vec{\chi}_l,\vec{\delta}_m$ ($k,l,m=0,1$)
denote the set of the phases defining the appropriate
observables. The measurement of each observable can yield three
possible results which we denote by $a$ for Alice, $b$ for Bob and
$c$ for Cecil ($a,b,c=1,2,3$). The probability
$P_{QM}(a_k,b_l,c_m)$, that Alice, Bob and Cecil obtain the
specific results after performing the unitary transformations
$U_A(\vec{\phi}_k)$, $U_B(\vec{\chi}_l)$ and $U_C(\vec{\delta}_m)$,
respectively, is given by the following formula
\begin{eqnarray}
&P_{QM}(a_k,b_l,c_m) =|\langle a_k|\langle b_l| \langle c_m|
U_A(\vec{\phi}_k)U_B(\vec{\chi}_l)
U_C(\vec{\delta}_m)|\psi\rangle|^2& \nonumber\\ &={1\over
27}+{1\over 27}\sum_{g'i'j'\neq gij}d_{g'i'j'}d_{gij} \nonumber
\\&
\times  \cos({2\pi\over 3} (a_k(g-g')+b_l(i-i')+c_m(j-j'))
+\phi_k^g-\phi_k^{g'}+\chi_l^i-\chi_l^{i'}+\delta_m^j-\delta_m^{j'})&,
\end{eqnarray}
where, for instance, $\phi_k^g$ denotes the $g$-th component of
$\vec{\phi}_k$.

%, and $U_X$, with $X=A,B,C$, are the unitary
%transformations associated with the action of the local device,
%endowed with the matrix elements of the form of (\ref{UNITARY}).

In the presence of random noise, in order to describe the system one has to
introduce the mixed state  $\rho_{F} =
(1-F)|\psi\rangle\langle\psi | + F\rho_{noise}$, where
$\rho_{noise}=\frac{1}{27}I$, and $I$ is the identity operator.
The non-negative parameter $F$ specifies the amount of noise
present in the system. In such a case, the quantum probabilities
read
$$P_{QM}^F(a_k,b_l,c_m) = (1-F)P_{QM}(a_k,b_l,c_m)+{F\over
27}$$.

The hypothesis of local realism assumes that there exists some
joint probability distribution $P_{LR}(a_0,a_1;b_0,b_1;c_0,c_1)$
that returns quantum probabilities $P_{QM}^F(a_k,b_l,c_m)$ as
marginals, e.g.,
\begin{eqnarray}
&&P_{QM}^F(a_0,b_0,c_0) \nonumber \\
&& =
\sum_{a_{1}=1}^3\sum_{b_{1}=1}^3\sum_{c_{1}=1}^3P_{LR}(a_0,a_1;b_0,b_1;c_0,c_1).
\label{marginals1}
\end{eqnarray}
Please note, that a concise notation of the full set of such
conditions can be given by
\begin{eqnarray}
&&P_{QM}^F(a_k,b_l,c_m) \nonumber \\
&& =
\sum_{a_{k+1}=1}^3\sum_{b_{l+1}=1}^3\sum_{c_{m+1}=1}^3P_{LR}(a_0,a_1;b_0,b_1;c_0,c_1).
\label{marginals}
\end{eqnarray}
where $k+1,l+1,m+1$ are understood as modulo $2$. For each pure
state $|\psi\rangle$, one can find the threshold $F_{thr}$ (the
{\em minimal}
value of $F$) above which such a joint probability
distribution satisfying (\ref{marginals}) exists (obviously, for
any separable state $F_{thr}=0$, however this may hold also for
some non-separable states).

There is a well defined mathematical procedure called linear
programming that allows us to find the threshold $F_{thr}$ for the
given state $|\psi\rangle$ and for the given set of observables.
We should stress that $F_{thr}$ found in this way gives us
sufficient and necessary conditions for violation of local
realism. The procedure works as follows.

The computation of the threshold $F_{thr}$ is equivalent to
finding the joint probability distribution
$P_{LR}(a_0,a_1;b_0,b_1;c_0,c_1)$, i.e., the set of $3^6$ of
positive numbers summing up to one and fulfilling $8 \times 27 =
216$ conditions given by (\ref{marginals}) such that $F$ is
minimal. Therefore, $F$ and $P_{LR}(a_0,a_1;b_0,b_1;c_0,c_1)$ can
be treated as variables lying in a $3^6 + 1$-dimensional real
space. The set of linear conditions (\ref{marginals}) and the
condition that $0 \leq F\leq 1$ defines a convex set in this
space.

Next, we define a linear function, whose domain is the convex set
defined above so that it returns the number $F$. The task of
finding $F_{thr}$ is then equivalent to the search for the minimum
of this function. As the domain of the function is very
complicated, the procedure can only be done numerically (we have
used the numerical procedure HOPDM 2.30, see \cite{gondzio}).

It is obvious that the $F_{thr}$ depends on the observables
measured by Alice, Bob and Cecil (which in turn depend on the set
of phases) as well as on the state $|\psi\rangle$ (indeed, for
some unfortunate choices of observables, or the states or both,
one can have $F_{thr}=0$). Let us clarify, that the task of the
linear optimization procedure is each time to find the {\em
minimal} $F$, for which the relation (\ref{marginals}) can be
satisfied  by some positive probabilities on its right hand side.
However, the left hand side of Eq. (\ref{marginals}) depends on
the chosen states and observables, and we are interested in the
case when getting the local realistic model requires a maximal
possible admixture of noise, therefore we search for such states
and observables, for which the minimal $F _{thr}$ has {\em the
largest possible value}. There are two possible interesting
scenarios. We can fix the state $|\psi\rangle$ and maximize
$F_{thr}$ over the observables. In this way we find the best
violation of local realism for this given state. Alternatively, we
can maximize $F_{thr}$ over the coefficients defining the state,
as well as over the observables. This procedure allows us to find
the optimal state, and optimal observables measured on this state,
which can yield the best possible violation of local realism by
the class of pure states with real coefficients (\ref{STATE}). Of
course, we do not have to limit ourselves to pure states with real
coefficients, nor even to pure states but then in these cases the
number of parameters over which we have to optimize becomes too
large for our computers to handle.

%\section{Calculations}

We have applied the procedure described above for the fixed state
$|\psi\rangle$, which we have chosen to be a maximally entangled
state, i.e., $|\psi\rangle = {1\over \sqrt 3}(|111\rangle +
|222\rangle+ |333\rangle$. Running the program we have found that
the threshold amount of noise, that has to be
admixed to the
maximally entangled state, so that the correlations generated by
it, for {\em any} sets of pairs of local settings of the phases,
become describable in a local and realistic way, is $F_{thr}
=0.4$. The optimal observables form the point of view of
violations of local realism, i.e., exactly those for which the
noise admixture must be maximal to get a local realistic model,
are defined by the following sets of phases $\vec{\phi}_0 =
(0,0,{2\over 3}\pi), \vec{\phi}_1=(0,0,0); \vec{\chi}_0=(0,0,\pi),
\vec{\chi}_1 = (0,0,{5\over 3}\pi); \vec{\delta}_0 = (0,{1\over
3}\pi,0),\vec{\delta}_1=(0,\pi,0)$. We can therefore say, that the
violation of local realism in this case is stronger than for two
maximally entangled qutrits, in which case the threshold amount of
noise is only $0.304$. However, it is weaker than the violation by
three entangled qubits, for which the threshold amount of noise is
$0.5$.

Naturally, one should check whether one can obtain better
violations for non maximally entangled states. Therefore we have
taken the predictions for (\ref{STATE}), and used a procedure for
the maximalization of $F_{thr}$ over the parameters $d_{gij}$ as
well as the observables.

We have found that the there exists a non-maximally entangled state,
and a certain set of local observables, for which one
requires  $F_{thr} = 0.571$ noise admixture for the correlations
to have a local realistic description. The expansion coefficients
of the state are given in the table below, whereas the phases
defining the optimal observables will not be presented here, as
they are not easily interpretable. However, for very close local
settings given by: $\vec{\phi}_0 = (0,{2 \over 3}\pi,-{5 \over
9}\pi), \vec{\phi}_1=(0,{2 \over 3}\pi,0); \vec{\chi}_0=(0,{17
\over 18}\pi, -{1 \over 18} \pi), \vec{\chi}_1 = (0,0,0);
\vec{\delta}_0 = (0,\pi,{23 \over 36} \pi), \vec{\delta}_1=(0,{7
\over 36}\pi,-{2 \over 3}\pi)$, there is a state for which the
threshold noise equals $0.570$.

\begin{widetext}
\begin{center}

\begin{tabular}{|l|rrrrrrrrr|}
\hline
Basis & \ket{000} & \ket{001} & \ket{002} & \ket{010} & \ket{011} & \ket{012} & \ket{020} & \ket{021} & \ket{022}\\ \hline
Coeff & $+0.186$  & $+0.076$  & $+0.230$  & $+0.218$  & $+0.046$  & $+0.112$  & $+0.172$  & $+0.033$  & $+0.247$ \\ \hline
\end{tabular}

\vspace{0.5cm}

\begin{tabular}{|l|rrrrrrrrr|}
\hline
Basis & \ket{100} & \ket{101} & \ket{102} & \ket{110} & \ket{111} & \ket{112} & \ket{120} & \ket{121} & \ket{122}\\ \hline
Coeff & $+0.216$  & $+0.050$  & $+0.110$  & $+0.160$  & $+0.049$  & $+0.236$  & $+0.204$  & $+0.055$  & $+0.235$ \\ \hline
\end{tabular}

\vspace{0.5cm}

\begin{tabular}{|l|rrrrrrrrr|}
\hline
Basis & \ket{200} & \ket{201} & \ket{202} & \ket{210} & \ket{211} & \ket{212} & \ket{220} & \ket{221} & \ket{222}\\ \hline
Coeff & $-0.078$  & $+0.406$  & $-0.029$  & $-0.023$  & $+0.385$  & $+0.035$  & $-0.123$  & $+0.393$  & $-0.128$ \\ \hline
\end{tabular}

\end{center}
\end{widetext}

In summary, we have shown, that for the maximally entangled state
three entangled qutrits violate local realism stronger than two
entangled qutrits (the threshold amount of noise $0.304$, see
\cite{KASZLIKOWSKI-PRL-2000}).  The threshold amount of noise to
get local realistic correlations is $0.4$. This violation is not
as strong as for three entangled qubits for which one has to admix
$50\%$ of noise to make the system describable by local realistic
theories. However, we can obtain much a stronger violation for the
non-maximally entangled states. In this case there exists a
non-maximally entangled state (see the table) for which $F_{thr} =
0.57$, i.e., we have to add $57\%$ of noise before we enter the
region in which the state admits local and realistic description.

We must stress, that although for the state given in the table,
the threshold amount of noise $F_{thr} = 0.57$ gives the necessary
and sufficient conditions for the existence of local realism, for
the measurement of the observables given by unbiased symmetric
three-port beamsplitters, it does not mean that with a different
choice of observables, or by allowing complex coefficients in the
state (\ref{STATE}), one cannot increase $F_{thr}$.

Moreover, it is reasonable to expect, that for four or higher
number of entangled qutrits the difference between the robustness
against noise (i.e., the resistance of quantum correlations to
classical description) of maximally entangled states and
non-maximally entangled ones will still increase.  Note that,
optimal non-maximally entangled state of two qutrits (for which
the threshold amount of noise is 0.3139) is around $3\%$ more
resistant to noise than the maximally entangled one (for which the
threshold amount of noise is 0.3038). In the case of three
entangled qutrits the difference between the threshold amount of
noise for non-maximally entangled state (0.571) and for maximally
entangled state (0.4) is about $40 \%$!

MZ thanks Nicolas Gisin for discussions on this topic. MZ and DK
acknowledge the support of KBN, project No. 5 PO3B 088 20. DK, LCK
and CHO would also like to acknowledge the support of A$\ast$Star
Grant No: 012-104-0040.

\end{document}